\documentclass[preprint,prd,aps,tighten,nofootinbib,amssymb]{revtex4}
\usepackage{graphicx}% Include figure files
\usepackage{dcolumn}% Align table columns on decimal point
\usepackage{bm}% bold math
\usepackage{amsmath}
\usepackage{color}

\input{colordvi.tex}
\newcommand{\CO}{{\cal O}}
\newcommand{\vect}[1]{\mbox{\boldmath${#1}$}}

\newcommand{\beq}{\begin{equation}}
\newcommand{\eeq}{\end{equation}}
\newcommand{\beqa}{\begin{eqnarray}}
\newcommand{\eeqa}{\end{eqnarray}}

\begin{document}

\title{Inflationary cosmology and the standard model Higgs with a small Hubble-induced mass}

\author{Kohei Kamada}
\email[Email: ]{kohei.kamada"at"epfl.ch}
\affiliation{ Institut de Th\'eorie des Ph\'enom\`enes Physiques,
\'Ecole Polytechnique F\'ed\'erale de Lausanne, 1015 Lausanne, Switzerland}


\begin{abstract}

We study the dynamics of the standard model Higgs field in the inflationary cosmology. 
Since metastability of our vacuum is indicated by the current experimental data of the Higgs boson and top quark, inflation models with a large Hubble parameter may have a problem: 
In such models, the Higgs field rolls down towards the unwanted true vacuum due to the large fluctuation in the inflationary background. However, this problem can be relaxed by supposing an additional mass term for the Higgs field generated during and after inflation. We point out that it does not have to be larger than the Hubble parameter if the number of $e$-folds during inflation is not too large. We demonstrate that a high reheating temperature is favored in such a relatively small mass case and it can be checked by future gravitational wave observations. Such an induced mass can be generated by, {\it e.g.,}  a direct coupling to the inflaton field or nonminimal coupling to gravity. 

\end{abstract}
\maketitle

\section{Introduction}

Recent results at the Large Hadron Collider (LHC) \cite{Aad:2012tfa,Chatrchyan:2012ufa} are in excellent agreement with the Standard Model (SM) with a 125 GeV Higgs boson and thus far 
any significant deviation from the SM has not been reported. On the other hand, the current measurements of the Higgs and top quark masses \cite{Lancaster:2011wr} suggest the metastability of our vacuum \cite{Espinosa:2007qp,EliasMiro:2011aa,Buttazzo:2013uya} (see also Ref.~\cite{Branchina:2013jra}); the Higgs potential becomes negative typically at $h \gtrsim 10^{11}$ GeV \cite{Buttazzo:2013uya}. It may be an important hint for high-energy physics.

One of the important ingredients in modern cosmology is inflation. It expands the primordial Universe at an accelerating rate. It solves the flatness and horizon problems and sows the seeds of the large scale structure of the present Universe. 
Within the current errors, there still remains a possibility of the SM-Higgs-driven inflation 
\cite{Bezrukov:2007ep}.  
However, if the Higgs potential is negative at $h \gtrsim 10^{11}$ GeV,
such Higgs inflation models cannot occur unless there is a physics beyond the SM that keeps the Higgs potential positive up to the inflationary scale  
because the Higgs field value during inflation is required to be larger than $10^{16-17}$ GeV
in these models.  
In this paper, we assume that the electroweak vacuum is metastable and inflation is driven by 
a scalar field other than the SM Higgs field, called inflaton.

The current data suggests that the lifetime of the electroweak vacuum is longer than the age of the Universe \cite{Arnold:1989cb}, and there is no constraint on the reheating temperature 
from the thermal-fluctuation-triggered electroweak vacuum decay \cite{Espinosa:2007qp,EliasMiro:2011aa,Arnold:1991cv}. However, the vacuum fluctuation in the quasi-de Sitter background of the Higgs field during inflation may also push it to the unwanted Anti de Sitter (AdS) vacuum if the Hubble parameter during inflation is large, {\it e.g.}, 
as the recent BICEP2 result suggests \cite{Ade:2014xna}.\footnote{
The recent result of Planck \cite{Adam:2014bub} suggests that the signals that 
BICEP2 observed may mainly come from the dust foreground. But one cannot conclude it at least before Planck $B$-mode results.}
Thus, it may spoil inflation or, at least, our Universe that lands in the metastable vacuum may be unlikely.\footnote{
Note that there are still discussions whether it is catastrophe for cosmology or not \cite{Espinosa:2007qp,Kobakhidze:2013tn,Hook:2014uia}.
}
Therefore, low-energy scale inflation may be favored in this viewpoint, %in which 
contrary to the BICEP2 result \cite{Ade:2014xna}, 
as discussed in other recent literatures \cite{Fairbairn:2014zia,Herranen:2014cua}.\footnote{
See also Ref.~\cite{Figueroa:2014aya} for the gravitational wave background
generated by the dynamics of the SM Higgs field after inflation.}

As pointed out in Refs.~\cite{Espinosa:2007qp,Lebedev:2012sy,Hook:2014uia}, it can be avoided by supposing a coupling between inflaton and the SM Higgs field without giving any major effects on the dynamics of inflaton.  
This is because the coupling produces the ``Hubble-induced mass'' during inflation, 
which pushes the field value where the Higgs potential goes negative to a much larger value.
In the case where the induced mass is much larger than the Hubble parameter \cite{Lebedev:2012sy} and the Higgs potential remains positive up to the Planck scale, the Higgs field is quickly pushed to the origin and its fluctuation is suppressed. Thus, the unwanted vacuum decay can be avoided even if the initial field value of the Higgs field is relatively large, $\sim 0.1 M_{\rm Pl}$ with $M_{\rm Pl}$ being the reduced Planck mass. Consequently, the electroweak vacuum can be naturally selected.

On the other hand, if the induced mass is smaller than the Hubble parameter, it seems to be difficult to suppress the quantum fluctuations and hence the vacuum decay cannot be avoided even if the Higgs field initially sits at the origin. In this paper, however, we point out that if the number of $e$-folds during inflation is not too large, we can construct a scenario with a high-scale inflation in which most part of the Universe can avoid the vacuum decay while the induced mass is not so large, as is also recently suggested in Ref.~\cite{Hook:2014uia}. 
This is because 
the evolution of the expectation value of the Higgs field during inflation is suppressed
and it can be than the field value of the potential barrier
if the Hubble-induced mass $m_H$ is large enough, $\Delta m_h^2/H_{\rm inf}^2\gtrsim 2 \times 10^{-2}$ and the number of $e$-folds during inflation is not too large. In addition, 
if the reheating temperature is high enough,  the present Universe can be safely realized.  
Note that after inflation the Higgs field still slow-rolls and the time-dependent potential barrier may 
catch it up. The Higgs field will roll down towards the unwanted AdS vacuum in this case. 
If the Higgs field is thermalized before being caught up by the potential barrier, the Higgs field safely settles down to the electroweak vacuum.
Owing to a relatively high reheating temperature, the Higgs field is thermalized earlier.  
Here we give a rough estimate for such a healthy scenario. 
We also point out that it would be possible to verify such a high reheating temperature  by the future gravitational wave experiments.

\section{Fluctuation of the Higgs field with a small induced mass during inflation \label{sec2}}

Let us start from the SM Higgs potential. At the large field values $h \gg v\equiv246$ GeV, it is well described by
\begin{equation}
V(h)=\frac{1}{4}\lambda(h) h^4, 
\end{equation}
in the unitary gauge. 
The Higgs quartic coupling $\lambda(h)$ runs logarithmically with respect to $h$ from $\lambda(M_h)\simeq 0.13$ 
where $M_h$ is the Higgs mass. 
As is studied in Ref.~\cite{Espinosa:2007qp,EliasMiro:2011aa,Chetyrkin:2012rz,Buttazzo:2013uya}, 
the Higgs quartic coupling becomes negative at $h \sim 10^{11}$ GeV. 
Though the uncertainties in the Higgs and top mass data lead to 
the uncertainty in the point where the potential goes negative  ranging 
from $10^9$ GeV to the Planck scale or higher, 
here we consider the case where the Higgs potential vanishes typically at  $10^{11}$ GeV. 
Then, the Higgs potential has also a maximum or a ``barrier'' at $h=\Lambda_0 \sim 10^{11}$ GeV. 
If the Hubble parameter during inflation\footnote{The subscript ``inf'' represents that 
the variable is evaluated at the inflationary era.} $H_{\rm inf}$ is larger than $\Lambda_0$, 
the fluctuation of the Higgs field easily climbs up the potential barrier and rolls 
down to the unwanted true vacuum during inflation 
even when it initially sits at the origin~\cite{Espinosa:2007qp,Kobakhidze:2013tn,Hook:2014uia,Fairbairn:2014zia}. 
It is claimed in Ref.~\cite{Espinosa:2007qp} that the regions or the bubbles where the Higgs field 
falls into the unwanted true vacuum collapse due to the AdS instability 
and hence only the regions where the Higgs field is inside the potential barrier may remain.  
Consequently the metastable electroweak vacuum and high-scale inflation may be 
compatible.\footnote{See also the discussion
in Ref.~\cite{Hook:2014uia}.} 
However, it is not clear whether the Universe expands properly by inflation and 
the AdS bubble does not cause any cosmological disasters. 
In particular, if the AdS bubbles of the true vacuum ``eat'' the region where the present electroweak 
vacuum is selected, the existence of our Universe falls into a crisis. 
Therefore, we can say inflation with a relatively small Hubble parameter 
$H_{\rm inf}< \Lambda_0$ is safe in the light of the current data of the Higgs and top mass.  
It is contradictory to the recent BICEP2 result, which suggests $H_{\rm inf} \simeq 10^{14}$ GeV \cite{Ade:2014xna}, if the observed $B$-mode is generated by the primordial gravitational waves.  

As is pointed out in Refs.~\cite{Espinosa:2007qp,Hook:2014uia} and studied in detail in Ref.~\cite{Lebedev:2012sy}, 
the Higgs field can acquire a Hubble-induced mass due to its interaction with the inflaton  $\phi$. 
For example, the ``Higgs-portal'' coupling
\begin{equation}
\Delta V=\frac{1}{2}\kappa \phi^2 h^2
\end{equation}
with $\kappa>0$ gives an effective positive mass squared $\kappa \langle \phi^2\rangle$ during and after inflation. 
Here the bracket represents the time average.  
In the case of massive chaotic inflation $V(\phi)=m^2 \phi^2/2$, we have 
$3H_{\rm inf}^2 M_{\rm Pl}^2=m^2 \phi_{\rm inf}^2/2$ during inflation and 
$3 \langle H^2 \rangle M_{\rm Pl}^2 = m^2 \langle \phi^2 \rangle$ in the inflaton oscillation dominated era after 
inflation.\footnote{Note that the kinetic energy and potential energy are equilibrated, $m^2 \langle \phi^2 \rangle/2 =\langle {\dot \phi}^2 \rangle/2$, at the oscillating phase. }
Thus, the effective Higgs mass squared is proportional to the Hubble squared both during and after inflation, 
$\Delta m_h^2\simeq \kappa (M_{\rm Pl}/m)^2 H^2$. 
Note that in order for the quantum correction not to dominate the tree level potential, $\kappa \lesssim 10^{-6}$ is required~\cite{Lebedev:2012sy}.

A similar effect can be achieved by a non-minimal coupling of the Higgs field to gravity.\footnote{
Such a coupling is also studied recently in Ref.~\cite{Herranen:2014cua}, where 
the running of the nonminimal coupling to gravity up to the electroweak 
scale is carefully studied. Since here we study the dynamics of the SM Higgs during and after inflation in detail, 
our study is complementary to Ref.~\cite{Herranen:2014cua}.}
Suppose that the Einstein-Hilbert action is replaced by 
\begin{equation}
\frac{\cal L}{\sqrt{-g}}=-\frac{1}{2}(M_{\rm Pl}^2+\xi h^2)R, 
\end{equation}
where $g$ is the determinant of the metric, 
$\xi$ is a negative parameter, 
and $R$ is the scalar curvature. 
The effect of this term can be seen easily in the Einstein frame. 
By performing the conformal transformation  and changing the frame to the Einstein frame, 
we get the effective Higgs potential as
\begin{equation}
\Delta V\simeq -\left( 2V(\phi) - \frac{{\dot \phi}^2}{2}  \right)\frac{\xi}{M_{\rm Pl}^2}h^2  \left(1+\CO\left(\frac{\xi h^2}{M_{\rm Pl}^2}\right)\right). 
\end{equation}
During inflation we have $3H^2 M_{\rm Pl}^2 \simeq V(\phi)$, and during
inflaton oscillation dominated era after inflation we have $3H^2 M_{\rm Pl}^2=V(\phi)+{\dot \phi}^2/2$ 
with $\langle V (\phi)\rangle \simeq \langle  {\dot \phi}^2 \rangle/2$. 
Here we assumed that the inflaton oscillates in the quadratic potential around its potential minimum. 
Thus, the Higgs field acquires positive mass squared $-\gamma \xi H^2$ during and after inflation 
with $\gamma$ being a parameter of order of $\CO(1-10)$.

Motivated by the interactions discussed above, now we consider a simple modification of the Higgs potential during inflation, 
\begin{equation}
\Delta V (h) = \frac{1}{2}c_{\rm inf} H_{\rm inf}^2 h^2 
\end{equation}
with $c_{\rm inf}$ being a positive numerical parameter. Here we consider the case $c_{\rm inf}\lesssim \CO(1)$ 
and study vacuum fluctuation in this potential. 
For $H_{\rm inf} \gg \Lambda_0$, the Hubble-induced potential overwhelms the original potential 
around $h \sim \Lambda_0$ and the potential barrier moves to a higher field value. 
In principle, we should calculate the running of the couplings to study the dynamics
of the Higgs field. 
However, they vary only logarithmically with respect to $h$ and hence we can treat them as constants, {\it e.g.,}
a negative quartic coupling $\lambda(h)={\tilde \lambda} \simeq -0.01$,  
in the first approximation. Then, we obtain the field value at the potential barrier as
\begin{equation}
\Lambda_h \simeq \sqrt{\frac{c_{\rm inf}}{-{\tilde \lambda}}} H_{\rm inf}, \label{lambdah}
\end{equation}
which is roughly ten times larger than the Hubble parameter during inflation for $c_{\rm inf}=\CO(1)$.

The Higgs field receives quantum fluctuations during inflation and acquires nonvanishing 
expectation value. If $c_{\rm inf}$ is not too small, we can neglect the quartic term in the potential  for the Higgs field.  
For the Higgs field that initially sits at the origin, the expectation value of the Higgs field is evaluated as \cite{Bunch:1978yq}
\begin{equation}
\langle h^2 \rangle_{\rm inf} =\frac{3H_{\rm inf}^2}{8 \pi^2 c_{\rm inf}}\left[1-\exp\left(-\frac{2c_{\rm inf}}{3}{\cal N}_*\right)\right], \label{condc}
\end{equation}
where ${\cal N}_*$ is the number of $e$-folds during inflation. 
It is still under discussion what is the correct survival condition,\footnote{
If the regions where experiences vacuum decay collapse into black holes and they evaporate quickly without destroying the stable electroweak vacua, vacuum decay during inflation 
is not dangerous (most optimistic case). 
On the other hand, if even only one region that experienced vacuum decay
takes over all the space and dominates the Universe, a vacuum decay in the past
light cone of the observable Universe causes catastrophe (most pessimistic case).  
In Ref.~\cite{Hook:2014uia}, the following discussion was held: 
In the former case, just a few more number of $e$-folds during inflation than the would-be 
number of $e$-folds is required to compensate the collapsed AdS regions and to 
reproduce our Universe. In the latter case, the Hubble-induced mass must be larger than 
$\delta m_h^2\gtrsim 0.5 H_{\rm inf}^2$. In Ref.~\cite{Herranen:2014cua}, the condition $V_{\rm max}^{1/4}>H_{\rm inf}$ is adopted for the stability condition. Here $V_{\rm max}$ is the potential energy at the potential barrier.   } and here we 
require $\langle h^2 \rangle < \Lambda_h^2$ as its representative.  
Then, we acquire the constraint on the Hubble-induced mass as 
\begin{equation}
c_{\rm inf}>\sqrt{\frac{-3 {\tilde \lambda}}{8\pi^2}}\simeq 1.9\times 10^{-2} \left(\frac{\tilde \lambda}{-0.01}\right)^{1/2}, \label{ccons}
\end{equation}
regardless of the Hubble parameter during inflation. 
Here we have approximated $1-\exp(-2c_{\rm inf}{\cal N}_*/3)\simeq 1$.\footnote{
The constraint Eq.~\eqref{ccons} is weaker than the one given in Ref.~\cite{Hook:2014uia}
since we are less pessimistic and we allow some vacuum decays in the past light cone
of the observable Universe. } 
Note that Eq.~\eqref{condc} neglects the quartic term in the potential, and hence 
at the boundary values of $c_{\rm inf}$ in Eq.~\eqref{ccons}, this approximation is no 
longer valid. The expectation value should be a little larger. 
However, the validity of this approximation recovers for a little larger value of $c_{\rm inf}$. 
In this sense, Eq.~\eqref{ccons} gives a most optimistic constraint that can be used 
as a reference. 

\begin{figure}[t]
 \begin{center}
\begin{tabular}{cc}
  \includegraphics[width=85mm]{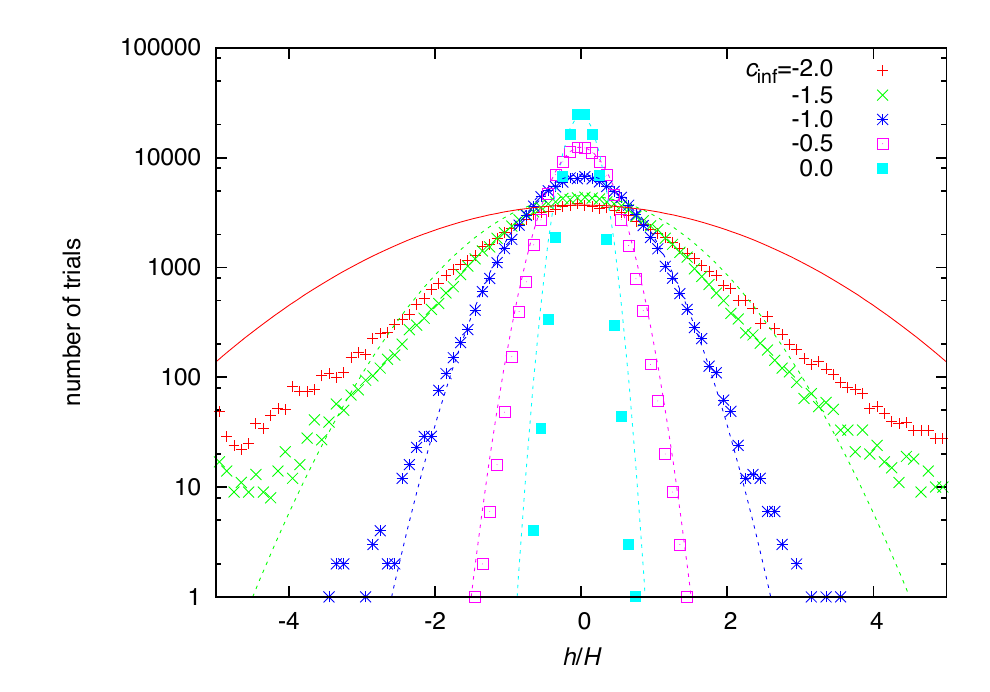}
  &
  \includegraphics[width=85mm]{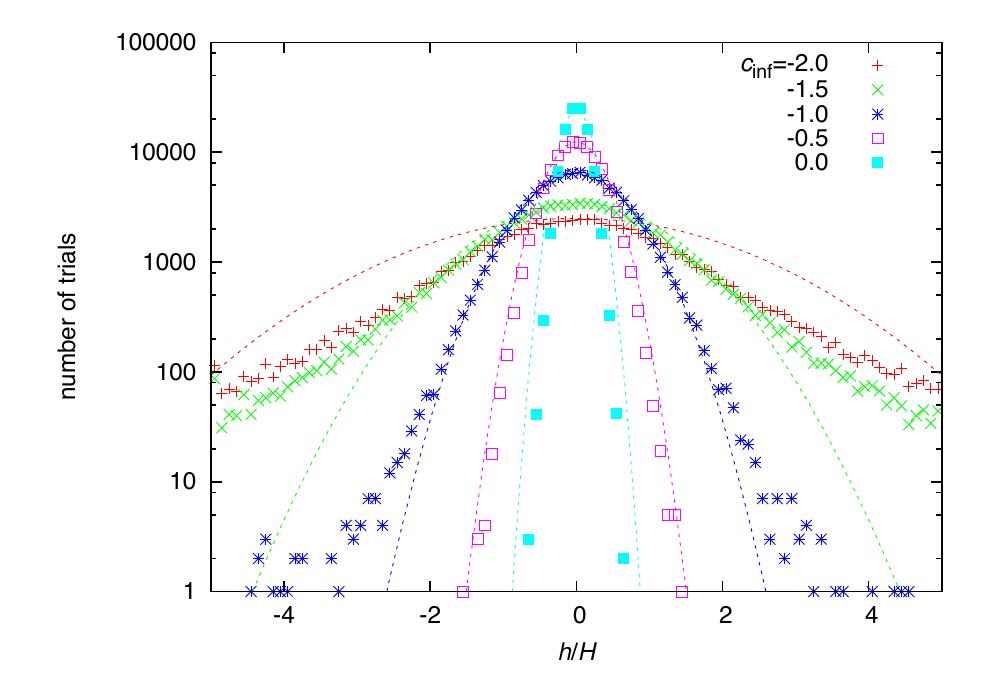} 
\end{tabular}
\vspace{-0.3cm}
 \end{center}
 \caption{The histogram of the Higgs field value at ${\cal N}_*=50$ (left) and 100 (right) with 
 $10^5$ trials. Dotted lines represent the Gaussian fitting, $\rho \propto \exp[-h^2/2\langle h^2 \rangle_{\rm inf}]$, with ${\cal N}_*\rightarrow \infty$ (Eq.~\eqref{condc}). 
 }
 \label{fig:hist} 
\end{figure}

\begin{figure}[t]
 \begin{center}
  \includegraphics[width=120mm]{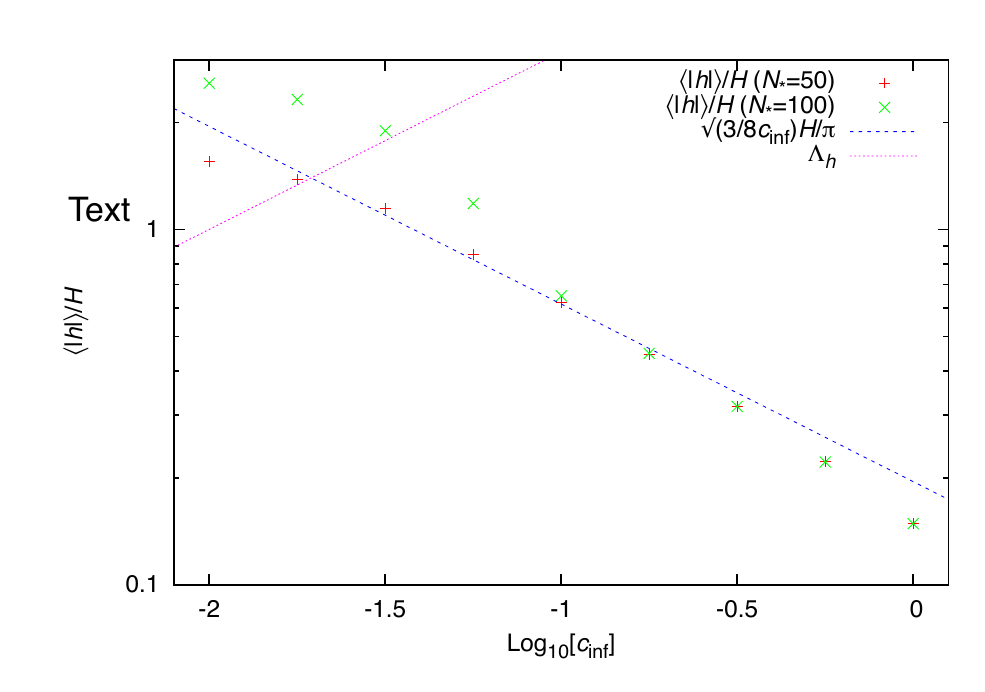}
\vspace{-0.3cm}
 \end{center}
 \caption{The numerical results of the 
 expectation values of the Higgs field at ${\cal N}_*=50$ and 100. 
 Blue dashed line represents Eq.~\eqref{condc} with ${\cal N}_*\rightarrow \infty$.  
 Purple dotted line represents $\Lambda_h$ (Eq.~\eqref{lambdah}).}
 \label{fig:ave} 
\end{figure}

Strictly speaking, we must calculate the probability distribution function (PDF) by using the stochastic approach 
or the Fokker-Planck approach
\cite{Stoch,Starobinsky:1994bd}
to estimate the survival probability. 
We have instead performed numerical calculation to solve the Langevin equations \cite{Sasaki:1987gy}. See Appendix \ref{appb} for the detail of the numerical calculation. 
Figure~\ref{fig:hist} shows the histogram of the Higgs field value at ${\cal N}_*=50$ and 100
for $c_{\rm inf}=10^{-2}, 10^{-1.5}, 10^{-1}, 10^{-0.5}, 1$ and ${\tilde \lambda}=-0.01$, with $10^5$ trials. 
We find that for $c_{\rm inf}>0.1$, the distribution is fitted by the 
Gaussian function with Eq.~\eqref{condc} (${\cal N}_*\rightarrow \infty$). 
One may surprised that the distribution is narrower than those expected Eq.~\eqref{condc} (${\cal N}_*\rightarrow \infty$) for smaller $c_{\rm inf}$.
This is because the distribution is during the 
course of (linear) spreading. As a result, many trials end inside the potential walls. 
Due to the negative quartic term, the tail of the distribution is broader than 
that of the Gaussian distribution and the Higgs field expectation value is not so small. 
Figure~\ref{fig:ave} shows the $c_{\rm inf}$ dependese of the Higgs field expectation value 
with ${\cal N}_*=50$ and 100. 
We can find for the ``just enough inflation'', ${\cal N}_*=50$, the expectation value of the 
Higgs field is well described by Eq.~\eqref{condc} with ${\cal N}_*=\infty$. 
%and find that the resultant 1-point PDF is well approximated by $\rho(h)\propto \exp[-h^2/2\langle h^2 \rangle_{\rm inf}]$ 
%if the above condition (Eq.~\eqref{ccons}) is satisfied. 
Therefore, we conclude that for the parameter that satisfies Eq.~\eqref{ccons}, the 
probability for the Higgs field to sit inside the potential barrier during inflation is not suppressed  exponentially and it 
gives an appropriately optimistic condition for the survival of the electroweak vacuum. 
Hereafter we use Eq.~\eqref{condc} with ${\cal N}_*=\infty$ 
as a representative constraint. 
We also use $c_{\rm inf}>10^{-1.5}$, which is the constraint for ${\cal N}_*=100$, 
as a reference.  
Note that for inflation with a larger number of $e$-folds, 
${\cal N}_*\gg 50$, the expectation value of 
the Higgs field diverges since the potential is not unbounded from the below, 
and the survival probability is, again, exponentially small. 
Therefore, the small Hubble-induced mass does not help the stability of the electro
weak vacuum during inflation. 
However, since for ${\cal N}_*\gg 50$, the expectation value we evaluated is the 
average in the whole Universe that is covered dominantly by unobservable region, 
anthropic principle would also matter, and hence we focus on the case 
where ${\cal N}_*\simeq 50.$
%One should note 
%that as explained above, this is just an approximate one and quantitatively 
%the real constraint should be a bit severer due to the inacuracy of Eq.~\eqref{condc}, 
%but the quantitative error would be small. 

%%%%%%%%%%%%%%%%%%%%%%%%%%%%%%%%

%%%%%%%%%%%%%%%%%%%%%%%%%%%%%%%%%%

%%%%%%%%%%%%%%%%%%%%%%%%%%%%%%%%

%%%%%%%%%%%%%%%%%%%%%%%%%%%%%%%%%%

\section{Dynamics of the Higgs field after inflation \label{sec:dyn}}

In the previous section, we give a(n approximate) condition in which the Higgs field does not 
roll down towards 
the unwanted true vacuum during inflation in many regions in space in the presence of the Hubble-induced mass. 
Is this condition a sufficient condition for us to live in the electroweak vacuum likely? 
The answer is no. 
Since the expectation value of the Higgs field just after inflation can be larger than the zero-temperature barrier $\Lambda_0$, 
we must consider the condition for the Higgs field to settle down to the electroweak vacuum 
through the dynamics after inflation. 

Let us consider a case where the Higgs field still receives a positive Hubble-induced mass 
during inflaton oscillation dominated phase,\footnote{The subscript ``osc'' represents that 
the parameter is evaluated at the inflaton oscillation dominated era.} 
\begin{equation}
\Delta V(h)=\frac{1}{2}c_{\rm osc} H(t)^2 h^2, \label{indosc}
\end{equation}
where $c_{\rm osc}\lesssim {\cal O}(1)$ is a numerical parameter. 
It would have a relation to $c_{\rm inf}$ but is model-dependent. 
Thus, we treat it as a different parameter. 
Note that the mechanism that induces the Hubble-induced mass discussed in the previous section does not exactly give 
the effective interaction described by Eq.~\eqref{indosc}, 
since it induces an oscillating term coming from the inflaton oscillation, 
say, $\phi(t)=\sqrt{2} {\bar \phi}(t) \cos mt$ where ${\bar \phi}$ is the slowly 
decreasing function and $m$ is the inflaton mass around its potential minimum. 
However, if the time scale of inflaton oscillation is much smaller than that of the Higgs field dynamics, 
the dynamics of the Higgs field and inflaton is decoupled and 
it is valid to take the time average of the oscillating part of inflaton.  
As a result,  Eq.~\eqref{indosc} gives the sufficiently well-approximated solution.
In our present case, the time scale of the Higgs field dynamics is given by $(c_{\rm osc}^{1/2}H(t))^{-1}$ 
whereas that of inflaton oscillation is given by $m^{-1}$. 
Since during oscillating stage the condition $m > H(t)$ is manifestly satisfied, 
it is safe to use the approximation Eq.~\eqref{indosc} 
for $c_{\rm osc}\lesssim {\cal O}(1)$ as an analytic estimation. 
In Appendix.~\ref{app}, we show the validity of this approximation by performing numerical calculation 
in a specific model. 

The Hubble parameter during inflaton oscillation dominated phase is well-approximated as
\begin{equation}
H(t)=\frac{2}{3t}.   \label{hubbleosc}
\end{equation}
This is the case when inflaton oscillates in the quadratic potential after inflation. 
During this phase before the complete reheating, partial decay of inflaton produces
relativistic particles as a subdominant component of the Universe. 
If their scattering cross section is large enough, they are thermalized with a temperature \cite{kolbturner}
\begin{equation}
T(t)=\left(\frac{72}{5 \pi^2 g_*(T)}\right)^{1/8} (H(t)M_{\rm Pl} T_R^2)^{1/4},\label{hubbletemp}
\end{equation}
where $g_*$ is the effective number of relativistic degrees of freedom, and $T_R$ is the reheating temperature. 
We here assume that 
at least the fields that do not have direct couplings to the Higgs such as gluons are 
thermalized just after inflation. 

If the reheating temperature is not high enough, the Higgs field is not thermalized just after 
the end of the inflation since the fields that coupled to the Higgs field acquire large mass which prevents 
them from thermalization, and hence Higgs field itself also cannot be thermalized.   
The Higgs field is thermalized when the particles that couples to Higgs field 
becomes light enough, $h(t)<T(t)$, and the interaction rate is rapid enough, 
$\Gamma \sim T(t)>H(t)$. Thus, in the case when the following conditions 
\begin{equation}
\langle h^2\rangle^{1/2}_{\rm inf} < T(t_{\rm inf})\Leftrightarrow T_R>\frac{3}{8\pi^2c_{\rm inf}} \left(\frac{5 \pi^2 g_*(T(t_{\rm inf}))}{72}\right)^{1/4} \frac{H_{\rm inf}^{3/2}}{M_{\rm Pl}^{1/2}} \equiv T_R^1, \label{tr1}
\end{equation}
and 
\begin{equation}
T(t_{\rm inf})>H_{\rm inf} \Leftrightarrow  T_R>\left(\frac{5 \pi^2 g_*(T(t_{\rm inf}))}{72}\right)^{1/4} \frac{H_{\rm inf}^{3/2}}{M_{\rm Pl}^{1/2}} \equiv T_R^2, 
\end{equation}
are satisfied, 
the Higgs field is thermalized just after the end of inflation. 
Otherwise, it takes some time for the Higgs field to be thermalized. 
(Or it is never thermalized as we will see.)  
One may wonder if the Higgs field is pushed to the unwanted AdS vacuum at the time of thermalization. 
It would be avoided if the Higgs field value is sufficiently small compared to the potential barrier 
$\Lambda_{\rm th} \simeq T(t_{\rm inf})/\sqrt{-\lambda}$ 
generated by the thermal potential $V_{\rm th} \simeq T^2 h^2$. 
From Eq.~\eqref{tr1}, we can easily see that if $T_R>T_R^1$, the Higgs field value just after inflation is roughly 
ten times smaller\footnote{Note that $\sqrt{-\lambda} \simeq 10^{-1}$.} than the potential barrier, 
which would be sufficiently small to avoid the disaster. 
Here we take $\langle h^2\rangle^{1/2}_{\rm inf}$ (Eq.~\eqref{condc}) with ${\cal N}_*\rightarrow \infty$ as a reference value of the Higgs field just after inflation. 
Note that for $c_{\rm inf}\simeq 0.02$, Eq.~\eqref{condc} is not precise and gives a lower 
bound of the expectation value as discussed, but the error is not so large as long as 
the number of $e$-folds is around 50. 
For larger values of $c_{\rm inf}$, the approximation Eq.~\eqref{condc} gets more precise.
Therefore, we will use $T_R^1$ and $T_R^2$ as references.   
We also do not write the coupling constants of the order of the unity, such as top Yukawa coupling, 
explicitly. 

Let us study the dynamics of the Higgs field before thermalization.  
The Higgs field evolves according to the potential
\begin{equation}
V(h)=\frac{1}{2}c_{\rm osc} H^2(t) h^2+\frac{1}{4}\lambda(h)h^4. 
\end{equation}
This potential has a time-varying maximum at 
\begin{equation}
h=\Lambda_t \simeq \sqrt{\frac{c_{\rm osc}}{-\lambda}}H(t), \label{potbaraftinf}
\end{equation}
for $\Lambda_t>\Lambda_0$ where $\lambda \simeq -{\cal O}(10^{-2})$ is negative. 
Thus, for the healthy realization of the present Universe, $h(t)<\Lambda_t$ must be satisfied in the course of the evolution of 
the Higgs field in substantial part of the Universe. Otherwise the Higgs field rolls 
down towards the unwanted AdS vacuum in many regions of the Universe, 
which may cause a cosmological disaster. 

Now we evaluate the Higgs field dynamics taking $\langle h^2 \rangle_{\rm inf}^{1/2}$ (Eq.~\eqref{condc}) with ${\cal N}_*\rightarrow \infty$ 
as the initial condition.\footnote{
Note once more that Eq.~\eqref{condc}  is not precise and just give an approximation for 
$c_{\rm inf}\sim 0.02$, though for our purpose the error is small enough, in particular 
for ${\cal N}_*\simeq 50$.  }
If the Higgs field does not roll down towards 
the unwanted AdS vacuum from this initial condition 
until its thermalization, 
the Higgs field successfully settles down to the electroweak vacuum in many regions of the Universe. 
Note that, again, there are two extreme possibilities. 
One is the possibility that the region where  the Higgs field 
rolls down to the AdS vacuum collapses to 
a black hole without destroying neighboring regions and 
evaporate quickly, any vacuum decay in the Universe is not problematic.
The other is  the one that even one vacuum decay in the past light cone of the observable 
Universe is dangerous 
if the bubble expands and takes over all the regions  of the Universe 
and it is dominated by the AdS vacua. 
Here, again, we instead give the survival condition as a condition that 
the Higgs field with the initial condition $\langle h^2_{\rm inf}\rangle^{1/2}$ 
does not roll down to the AdS vacuum. 
In principle, it would be better to perform a lattice simulation taking into account the spatial distribution 
of the Higgs fields. However, the spatial derivative of the Higgs field is suppressed due to inflation, and hence
we here only consider the homogeneously distributed Higgs field.\footnote{
We emphasize that since at later epoch causally disconnected region enter 
inside the horizon and hence gradient term may become important. 
Therefore, the results should be taken as only approximate ones and the gradient 
term may change the result slightly. } 
Then, the equation of motion (EOM) is given by
\begin{equation}
{\ddot h}(t)+3H(t){\dot h}+c_{\rm osc} H^2(t) h(t) + \lambda(h) h^3(t)=0. \label{eomeom}
\end{equation}
As long as $h(t)<\Lambda_t$ is satisfied, we can neglect the last term in the EOM 
and get a solution, 
\begin{equation}
h(t)=\langle h^2 \rangle^{1/2}_{\rm inf} \left(\frac{H(t)}{H_{\rm inf}}\right)^{(1-\sqrt{1-16c_{\rm osc}/9})/2}\simeq \sqrt{\frac{3}{2c_{\rm inf}}}\frac{H_{\rm inf}}{2\pi} \left(\frac{H(t)}{H_{\rm inf}}\right)^{(1-\sqrt{1-16c_{\rm osc}/9})/2}.  \label{higgsaftinf}
\end{equation}
Here we consider the case where $c_{\rm osc}<9/16$ and the Higgs field does not oscillate. 
Since the Higgs field value decreases slower than the potential barrier, $\Lambda_t$, 
we must seek for the way to avoid for the Higgs field to be caught up by the potential barrier 
after inflation for the successful Universe. Otherwise it rolls down to the unwanted AdS vacuum. 
This catching up would happen when 
\begin{equation}
h(t) \simeq \Lambda_t \Leftrightarrow H(t)\simeq \left(\frac{-3 \lambda}{8 \pi^2 c_{\rm osc} c_{\rm inf}}\right)^{1/(1+\sqrt{1-16 c_{\rm osc}/9})} H_{\rm inf}\equiv H_c. 
\end{equation}
Here we assumed that the approximations Eqs.~\eqref{potbaraftinf} and \eqref{higgsaftinf} hold until that time. 

The first way to avoid the falling down to the unwanted AdS vacuum is that the Higgs field gets thermalized 
before being caught up. 
Let us take the criteria for the Higgs field thermalization as $T(t)>h(t)$ and $T(t)>H(t)$.\footnote{
Note that in reality, thermalization does not completes instantaneously 
and completes a little later time than the one estimated in the below. 
This effect can be absorbed by the numerical factor $\alpha$ with the order of unity. 
See also Refs.~\cite{Espinosa:2007qp,EliasMiro:2011aa,Arnold:1991cv} for the potentially dangerous thermal-fluctuation-triggered electroweak vacuum decay. }
 
Then, we have two cases for the successful Universe when the Higgs field is not thermalized 
just after inflation;\footnote{In the case with $T_R<T_R^1$ and $T_R>T_R^2$, 
$h(t)<T(t)$ will never be satisfied because $T(t)$ decreases much  more rapidly than $h(t)$ 
in this parameter region. }
\begin{enumerate}
\item The case with $T_R>T_R^1$ and $T_R<T_R^2$:  
$H(t)=T(t)$ gets satisfied at a later time. 
For the successful Universe, both the conditions $T(t)>h(t)$ and $\Lambda_t>\alpha h(t)$ should be  
satisfied at $H(t)=T(t)$. 
\item The case with $T_R<T_R^1$ and $T_R<T_R^2$: For the successful Universe, the conditions $T(t)>h(t),  H(t)$ 
should be simultaneously satisfied before it gets $\Lambda_t<\alpha h(t)$. Note that it must be also satisfied before reheating 
because there are no longer ``Hubble-induced mass'' after inflaton decay. 
\end{enumerate}
One may wonder, again, if at the time when the Higgs field gets thermalized, thermal fluctuations may push the 
Higgs field to the unwanted AdS vacuum.   
At present we do not know how to calculate exactly the tunneling rate of a slow-rolling scalar field 
during the epoch when the system gets thermalized, 
unlike the case discussed in Ref.~\cite{Arnold:1991cv} where the Higgs field is at the potential minimum 
and the system is well-approximated to be in zero-temperature or fully thermalized. 
However, we may be allowed to guess it will be exponentially suppressed  
by using triangle approximation \cite{Duncan:1992ai} if the 
scalar field value is far away enough from the potential barrier and it is high enough. 
In this reason, we introduced a numerical parameter $\alpha\gtrsim \CO(1)$ in order to take into account it, 
though a careful study would be required to determine its value exactly, strictly speaking, but it is beyond the 
scope of this paper. 

{\bf Case 1}: $H(t)=T(t)$ is satisfied when
\begin{equation}
H(t)=\left(\frac{72}{5 \pi^2 g_*}\right)^{1/6}(M_{\rm Pl} T_R^2)^{1/3} \equiv H_{T1}.
\end{equation}
Then, the conditions $T(H_{T1})>h(H_{T1})$ and $\Lambda_t(H_{T1})>\alpha h(H_{T1})$ are rewritten
in terms of the constraint on the reheating temperature as
\begin{align}
T_R&>\left(\frac{5 \pi^2 g_*}{72}\right)^{1/4} \left(\frac{3}{8\pi^2 c_{\rm inf}}\right)^{\frac{3}{2(1+\sqrt{1-16 c_{\rm osc}/9})}} \frac{H_{\rm inf}^{3/2}}{M_{\rm Pl}^{1/2}} \equiv T_R^3, \\
T_R&>\left(\frac{5 \pi^2 g_*}{72}\right)^{1/4} \left(\frac{-3\alpha^2 \lambda}{8\pi^2 c_{\rm inf}c_{\rm osc}}\right)^{\frac{3}{2(1+\sqrt{1-16 c_{\rm osc}/9})}} \frac{H_{\rm inf}^{3/2}}{M_{\rm Pl}^{1/2}} \equiv T_R^4. 
\end{align}

{\bf Case 2}: In this case, $H(t)=T(t)$ is satisfied at $H=H_{T1}$ and $h(t)=T(t)$ is satisfied when 
\begin{equation}
H(t)=\left[2 \pi\sqrt{\frac{2 c_{\rm inf}}{3}} \left(\frac{72}{5 \pi^2 g_*}\right)^{1/8} \left(\frac{M_{\rm Pl} T_R^2}{H_{\rm inf}^3}\right)^{1/4} \right]^{4/(1-2\sqrt{1-16 c_{\rm osc}/9})} H_{\rm inf}\equiv H_{T2},  
\end{equation}
for  $c_{\rm osc}>27/64$.\footnote{In the case $c_{\rm osc}<27/64$, $h(t)$ will never catch up $T(t)$. } 
Then, we find that the rolling down problem is avoided if 
the Hubble parameter becomes $H_{T2}$ before the catching up time 
and reheating  in the parameters we are interested in. In other words, the present Universe will be realized if 
both the conditions are satisfied,  
\begin{align}
\Lambda_t&(H_{T2})>\alpha h(H_{T2}) \notag \\
 \Leftrightarrow &T_R>\frac{3}{8\pi^2 c_{\rm inf}}\left(\frac{5 \pi^2 g_*}{72}\right)^{1/4}\left(\frac{-3 \lambda}{8\pi^2 c_{\rm inf}c_{\rm osc}}\right)^{\frac{1-2\sqrt{1-16 c_{\rm osc}/9}}{2(1+\sqrt{1-16c_{\rm osc}/9})}}  \alpha^{\frac{1-2\sqrt{1-16 c_{\rm osc}/9}}{1+\sqrt{1-16 c_{\rm osc}/9}}}\frac{H_{\rm inf}^{3/2}}{M_{\rm Pl}^{1/2}} \equiv T_R^5, 
\end{align}
and 
\begin{align}
H_{T2}>H_R=&\left(\frac{\pi^2 g_*}{90}\right)^{1/2} \frac{T_R^2}{M_{\rm Pl}} \notag \\
\Leftrightarrow T_R>&\left(\frac{\pi^2 g_*}{90}\right)^{\frac{1-2\sqrt{1-16 c_{\rm osc}/9}}{8\sqrt{1-16c_{\rm osc}/9}}}\left(\frac{1}{2\pi}\sqrt{\frac{3}{2 c_{\rm inf}}}\left(\frac{5 \pi^2 g_*}{72}\right)^{1/8}\right)^{1/\sqrt{1-16 c_{\rm osc}/9}} \notag \\
 &\times H_{\rm inf}^{\frac{1+\sqrt{1-16 c_{\rm osc}/9}}{2 \sqrt{1-16 c_{\rm osc}/9}}} M_{\rm Pl}^{\frac{-1+\sqrt{1-16 c_{\rm osc}/9}}{2 \sqrt{1-16 c_{\rm osc}/9}}} \equiv T_R^6. 
\end{align}

In summary, if one of the following conditions, 
\begin{itemize}
\item $T_R>T_{R}^1$, and $T_{R}^2$
\item  $T_R<T_{R}^2$, and $T_R>T_{R}^1,T_{R}^3,T_{R}^4$
\item $T_R<T_{R}^1,T_{R}^2$, and $T_R>T_{R}^5,T_{R}^6$
\end{itemize}
are satisfied, the Higgs fields are thermalized before being caught up by the potential barrier 
and the present electroweak vacuum would be successfully selected.

The second way for the successful cosmic history is that the Higgs field value $h(t)$ becomes smaller 
than the zero-temperature barrier $\Lambda_0$ and gradually its dynamics is dominated by 
$\lambda h^4/4$ term before being caught up by the potential barrier. 
The Higgs expectation value $h(t)$ gets smaller than $\Lambda_0$ when 
\begin{equation}
H(t)<\left(\sqrt{\frac{8\pi^2 c_{\rm inf}}{3}}\frac{\Lambda_0}{H_{\rm inf}}\right)^{2/(1-\sqrt{1-16 c_{\rm osc}/9})} H_{\rm inf} \equiv H_\Lambda. 
\end{equation}
Thus, if $H_\Lambda>H_c$,  the present electroweak vacuum is successfully selected. 
This condition gives a constraint on the Hubble parameter during inflation as 
\begin{equation}
H_{\rm inf}<\left(\frac{8\pi^2 c_{\rm inf}}{3}\right)^{1/2} \left(\frac{-3\lambda}{8 \pi^2 c_{\rm inf} c_{\rm osc}}\right)^{\frac{1+\sqrt{1-16 c_{\rm osc}/9}}{1-\sqrt{1-16c_{\rm osc}/9}}} \Lambda_0. 
\end{equation}

\begin{figure}[t]
 \begin{center}
\begin{tabular}{cc}
  \includegraphics[width=85mm]{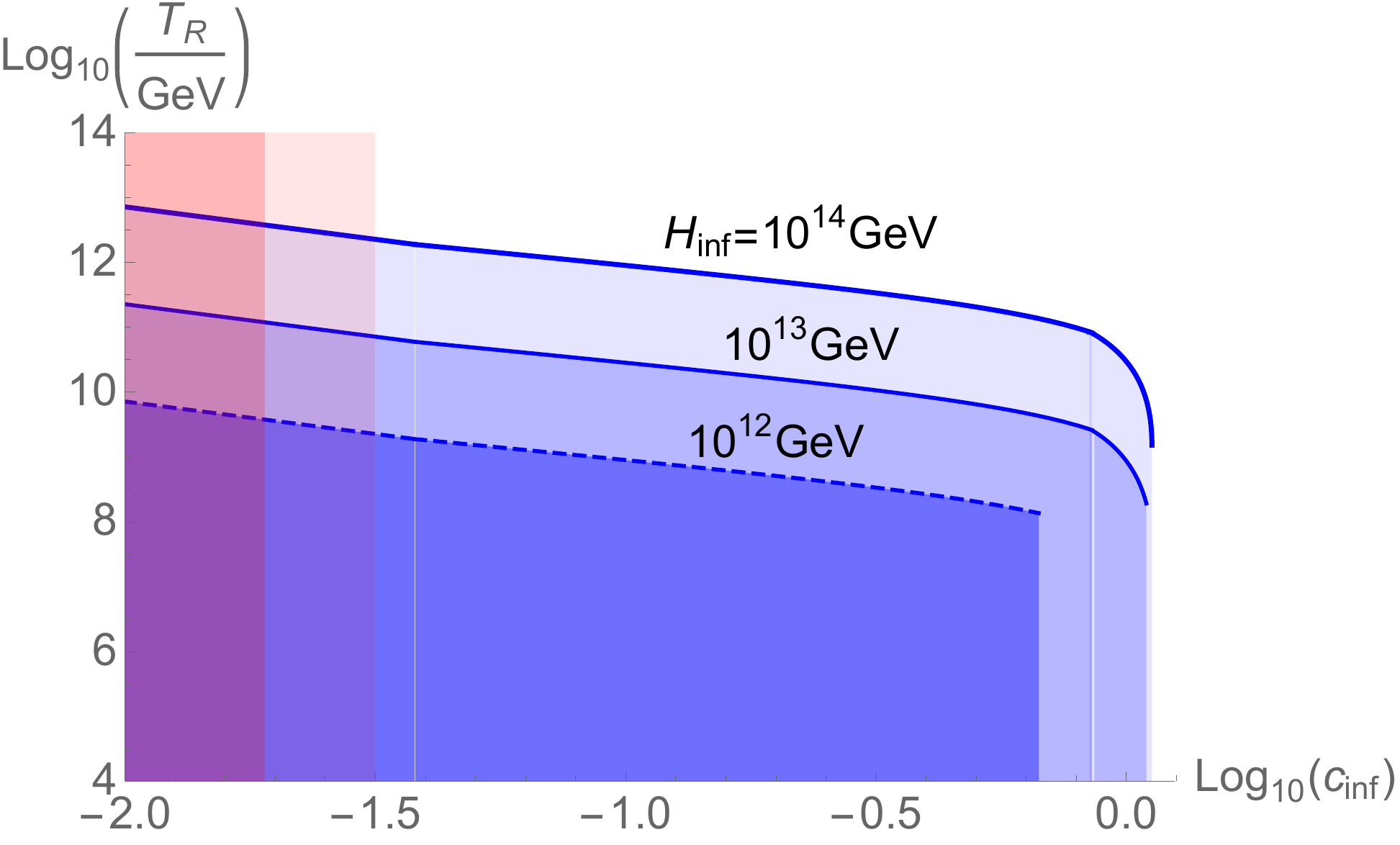}
  &
  \includegraphics[width=85mm]{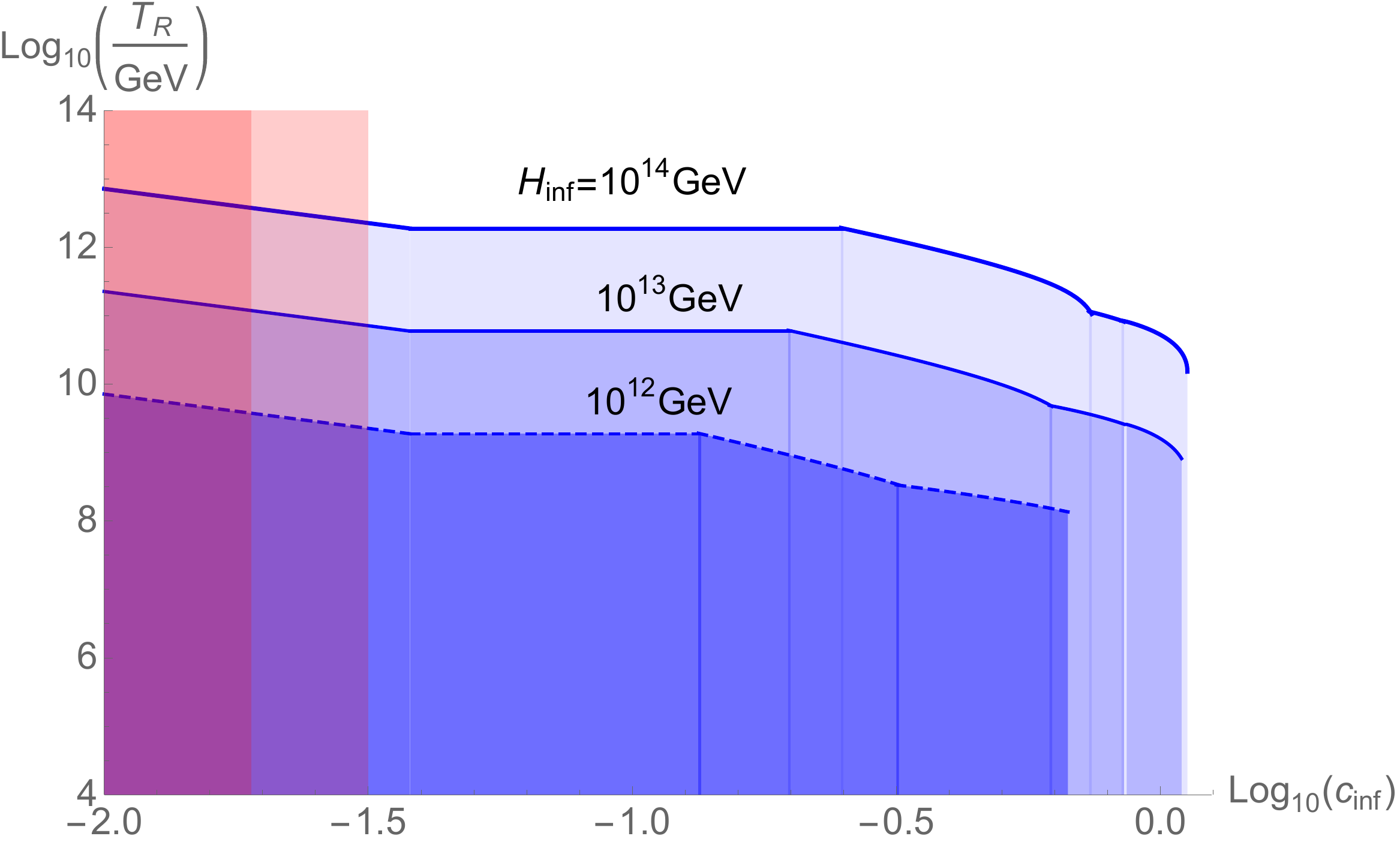} 
\end{tabular}
\vspace{-0.3cm}
 \end{center}
 \caption{The constraint on the reheating temperature according to the coupling constant $c_{\rm inf}$ with $c_{\rm osc}= c_{\rm inf}/2$. 
 The parameter $\alpha$ is chosen as 1.0 (left) and 10.0 (right). 
 }
 \label{fig:1} 
\end{figure}

\begin{figure}[t]
 \begin{center}
\begin{tabular}{cc}
  \includegraphics[width=85mm]{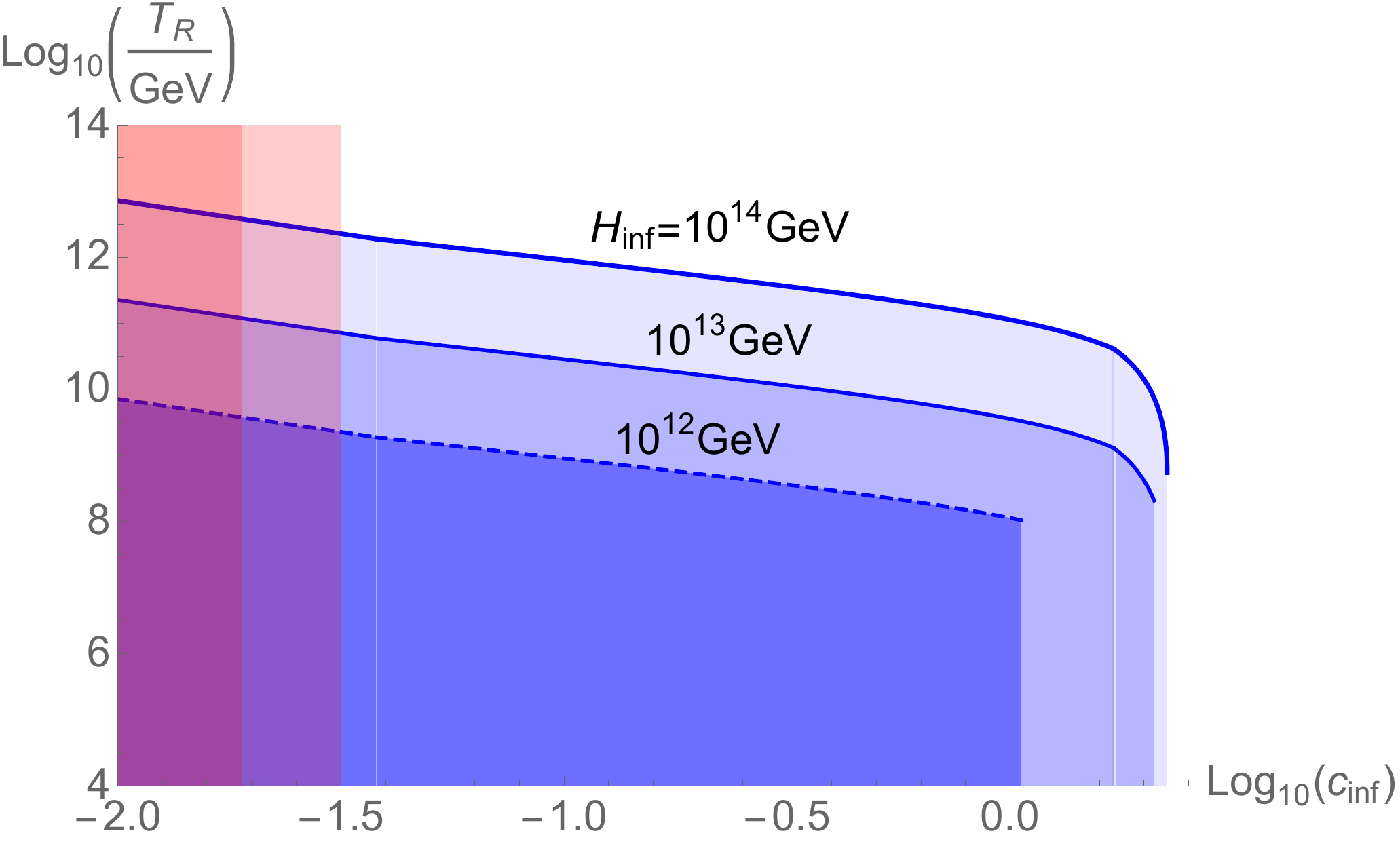}
  &
  \includegraphics[width=85mm]{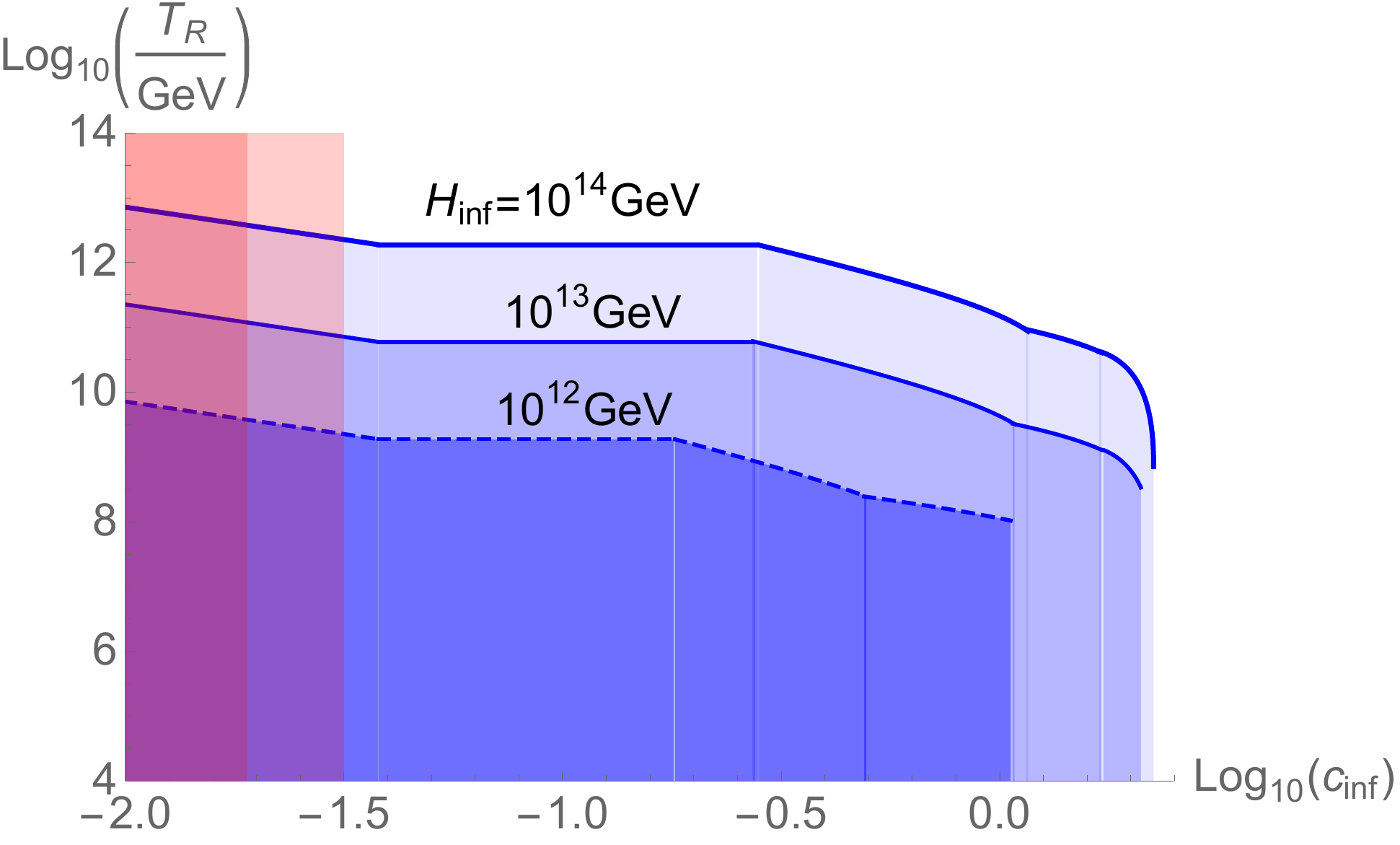} 
\end{tabular}
\vspace{-0.3cm}
 \end{center}
 \caption{The same to Fig. \ref{fig:1} but $c_{\rm osc}=c_{\rm inf}/4$.
 }
 \label{fig:2} 
\end{figure}

As an example, we show the allowed region in the $T_R$-$c_{\rm inf}$ plane in 
Figs.~\ref{fig:1} and \ref{fig:2}  for $H_{\rm inf}=10^{12}, 
10^{13}$ and $10^{14}$ GeV, respectively, with the parameters being chosen as 
$\Lambda_0=10^{11}$ GeV, and  $c_{\rm osc}=c_{\rm inf}/2$ (Fig.~\ref{fig:1}),  $c_{\rm inf}/4$ (Fig.~\ref{fig:2}).  
We also chose the value of  $\alpha$ as $\alpha=1.0, 10.0$. 
We find the constraint in analytic expression for this case as
\begin{equation}
T_R>\left\{
\begin{array}{ll}
T_R^1, & {\rm for} \quad c_{\rm inf}<\dfrac{3}{8\pi^2} \\ \\
{\rm min}.\{T_R^2, {\rm max}.\{ T_R^3, T_R^4 \} \},& {\rm for} \quad \dfrac{3}{8\pi^2}<c_{\rm inf}<\dfrac{27}{64} \times \dfrac{c_{\rm inf}}{c_{\rm osc}} \\ \\
T_R^5. & {\rm for} \quad \dfrac{27}{64}\times \dfrac{c_{\rm inf}}{c_{\rm osc}}<c_{\rm inf}<\dfrac{9}{16} \times \dfrac{c_{\rm inf}}{c_{\rm osc}}
\end{array}\right.
\end{equation}
We approximate the running of the Higgs quartic coupling  near $\mu\simeq 10^{11}$ GeV as \cite{Buttazzo:2013uya}
\begin{equation}
\lambda(\mu)=-1.4 \times 10^{-3}\ln \left(\frac{\mu}{10^{11} {\rm GeV}}\right) , 
\end{equation}
and evaluate it at $\mu=h(H_T)$ for $T_R=10^{11.5}, 10^{10}$, and $10^{8.5}$ GeV for $H_{\rm inf}=10^{14}, 10^{13}$, 
and $10^{12}$ GeV, respectively,  
when we give the lower bounds on $T_R$. 
The thick red colored region is disfavored due to the condition $\langle h^2 \rangle_{\rm inf}> \Lambda_h^2$ with Eq.~\eqref{condc} (${\cal N}_*\rightarrow \infty$), 
which means the survival probability during inflation is exponentially suppressed.\footnote{
Note that Eq.~\eqref{condc} is only an approximate expression, and 
this constraint is an optimistic constraint and should be regarded as a reference. 
But for ${\cal N}_*\simeq 50$, this approximation is accurate enough for our purpose 
as explained in Sec.~\ref{sec2}.  } 
We also show the constraint $c_{\rm inf}<10^{-1.5}$ that represents 
$\langle h^2\rangle_{\rm inf}<\Lambda_h^2$ for ${\cal N}_*\simeq 100$
in light red region. 
The blue colored regions are excluded due to the condition for the Higgs field not to fall into the unwanted
true vacuum. Note that there are no constraint for $c_{\rm inf}>10^{-0.17}(10^{0.04}) \ (c_{\rm osc}=c_{\rm inf}/2)$ and $c_{\rm inf}>10^{0.02}(10^{0.32}) \ (c_{\rm osc}=c_{\rm inf}/4)$ in the cases  $H_{\rm inf}=10^{12(13)}$ GeV, 
in which  the condition $h(t)<\Lambda_0$ is always satisfied before the rolling down to the unwanted true vacuum. 
The running of the quartic coupling $\lambda$ is not calculated strictly, 
but it does not change the result so much. 
We can see that the lower bound of the reheating temperature becomes severer as the Hubble parameter 
during inflation is larger. 
Increasing the parameter $\alpha$ makes the lower bound slightly higher, but it does not change the feature significantly. 
For smaller values of $c_{\rm osc}/c_{\rm inf}$, the excluded region is slightly enhanced at larger $c_{\rm inf}$, but the overall
feature does not change.   
For the Hubble parameter $H_{\rm inf}\simeq 10^{14}$ GeV, which is suggested by the recent BICEP2 result, 
a relatively high reheating temperature, $T_R \gtrsim 10^{12-13}$ GeV is required. 
This indicates that if the B-mode in the CMB polarization observation with $r\simeq 0.2$ is confirmed, 
the stochastic GW background {\it must} be detected in the gravitational detectors \cite{Nakayama:2008ip} 
such as DECIGO \cite{DECIGO} or BBO \cite{BBO} due to the relatively large reheating temperature. 
If not, it suggests that there is a physics beyond the SM to stabilize the Higgs potential \cite{Lebedev:2012zw}
or the Hubble-induced mass for the Higgs field during inflation is much larger than the Hubble parameter 
as is the case studied in Ref.~\cite{Lebedev:2012sy}. 

Note that Figs.~\ref{fig:1} and \ref{fig:2} assume the approximate expression
of the expectation value of the Higgs field during inflation Eq.~\eqref{condc} with 
${\cal N}_*\rightarrow \infty$. 
As explained in Sec.~\ref{sec2}, this expression is not accurate around
$c_{\rm inf}\simeq 0.02$, and the figures should be understood as approximate
estimates, in particular, for small values of $c_{\rm inf}$. 
But for ${\cal N}_*=50$, this approximation is valid enough for our purpose. 
Note also that we here do not take into account the spatial derivative in this study. 
This would decrease the lower bound of the reheating temperature 
since it makes the Higgs field damp more rapidly. 
It would be effective for larger $c_{\rm inf}$. 
For smaller $c_{\rm inf}$, however, our result suggests that the Higgs field 
must be thermalized just after inflation, before the spatial derivative would get effective, 
and hence the constraint will not change significantly. 
In summary, Figs.~\ref{fig:1} and \ref{fig:2}  are just approximate estimates. 
For smaller values of $c_{\rm inf}$, the expression Eq.~\eqref{condc} will induce slight errors and 
for larger values of $c_{\rm inf}$, there are small errors from 
the neglect of the  gradient terms in Eq.~\eqref{eomeom}. 
But qualitatively, these figures give us approximately correct constraints.

%%%%%%%%%%%%%%%%%%%%%%%%%%%%%%%%

%%%%%%%%%%%%%%%%%%%%%%%%%%%%%%%%%%
%%%%%%%%%%%%%%%%%%%%%%%%%%%%%%%%

%%%%%%%%%%%%%%%%%%%%%%%%%%%%%%%%%%

\section{Summary}

In this article, we studied the evolution of the SM Higgs field in the inflationary cosmology 
in the light of recent collider experiments, which suggests the metastability of the electroweak vacuum. 
If the electroweak vacuum is metastable, 
high-scale inflation may be problematic since the Higgs field rolls down to the 
unwanted AdS vacuum and the probability for the Higgs field to remain the electroweak vacuum 
is exponentially suppressed, though it is still under discussion if it is a real catastrophe for our Universe or not. 
We found that the Hubble-induced mass can avoid the exponentially 
suppressed survival probability of the electroweak vacuum during inflation while it is not necessarily
larger than the Hubble parameter during inflation if the number of $e$-folds during inflation 
is not too large. 
We also found that  the present Universe can be successfully 
realized even in the case of the relatively small Hubble-induced mass if the reheating temperature is high enough. 
This is because the Higgs field is thermalized before being caught up by the time-dependent 
potential barrier and before rolling down to the unwanted AdS vacuum. 
As a result, relatively high-energy scale inflation is allowed, 
and hence we can expect for the detection of GW background in the future experiments.  %without 
%any physics beyond the SM that is introduced to stabilize the Higgs potential. 
We also pointed out that the direct GW background detection will give us the clue to study the physics beyond the SM. 
Note that since the Higgs mass during inflation can be smaller than the Hubble parameter, 
it may be possible to generate a feature in the CMB, for example, nongaussianity, 
though it will require nontrivial interaction for the Higgs field.

\section*{Acknowledgments}
The author is grateful to M.~Asano, O.~Lebedev, and A.~Westphal for collaboration at the early stage of this project. 
The author also thanks M.~Shaposhnikov for useful comments. 
This work has been supported in part by the JSPS Postdoctoral Fellowships for Research Abroad. 

\appendix

\section{Numerical approach to the Langevin equation for the Higgs field during inflation \label{appb}}
Here we explain the numerical method we adopt to solve 
Langevin equation in Sec.~\ref{sec2}. 
The Langevin equation we here solve is \cite{Sasaki:1987gy}
\begin{align}
{\dot{\bar \phi}}(\vect{x},t) &= {\bar \pi} (\vect{x},t)+ \sigma(\vect{x},t), \label{Lange3}\\
{\dot{\bar \pi}}(\vect{x},t) &= -3 H {\bar \pi}(\vect{x},t)-\left.\frac{\partial V}{\partial \phi}\right|_{\phi={\bar \phi}}+\tau (\vect{x},t) , \label{Lange4}
\end{align} 
with correlation functions
\begin{align}
\langle 0 | \sigma(x_1) \sigma(x_2) | 0 \rangle =&\frac{\Gamma(\nu)^2}{2^{1-2\nu}}\frac{H^3}{4 \pi^3} \delta(t_1-t_2), \label{cor1}\\
\langle 0 | \tau(x_1) \tau(x_2) | 0 \rangle =& \frac{\Gamma(\nu)^2}{2^{1-2\nu}} \left|\nu-\frac{3}{2}\right|^2 \frac{H^5}{4 \pi^3} \delta(t_1-t_2),\label{cor2}\\
\frac{1}{2}\langle 0 | \sigma(x_1)\tau(x_2)+\tau(x_1) \sigma(x_2) | 0 \rangle =&\frac{\Gamma(\nu)^2}{2^{1-2\nu}} \left(\nu-\frac{3}{2}\right)\frac{H^3}{4 \pi^4} \delta(t_1-t_2). \label{cor4}
\end{align}
Here ${\bar \phi}$ is the corse-grained Higgs field and ${\bar \pi}$ is its canonical conjugate momentum. 
$\sigma$ and $\tau$ are stochastic noise terms. 
Redefining the field as
\begin{align}
{\tilde \phi} \equiv {\bar \phi}-\frac{1}{H(\nu-3/2)} {\bar \pi}, 
\end{align}
the equation of motion is rewritten as
\begin{align}
{\dot{\tilde \phi}}(\vect{x},t) &= \left(1+\frac{3}{\nu-3/2}\right) {\bar \pi} (\vect{x},t)+ \frac{1}{H(\nu-3/2)}\left.\frac{\partial V}{\partial \phi}\right|_{\phi = {\tilde \phi}+\pi/(H(\nu-3/2))}, \label{Lange3}\\
{\dot{\bar \pi}}(\vect{x},t) &= -3 H {\bar \pi}(\vect{x},t)-\left.\frac{\partial V}{\partial \phi}\right|_{\phi={\tilde \phi}+\pi/(H(\nu-3/2))}+\tau (\vect{x},t) , \label{Lange}
\end{align} 
with correlation function, 
\begin{equation}
\langle 0 | \tau(x_1) \tau(x_2) | 0 \rangle = \frac{\Gamma(\nu)^2}{2^{1-2\nu}} \left|\nu-\frac{3}{2}\right|^2 \frac{H^5}{4 \pi^3} \delta(t_1-t_2). 
\end{equation}
Note that the correlation function vanishes for the stochastic force for ${\tilde  \phi}$. 

To solve the Langevin equation for our system with 
$V=c_{\rm inf} H^2 {\bar \phi}^2/2+\lambda {\bar \phi}^4/4$, numerically, 
we normalize the time and field values with respect to the Hubble parameter; 
$N\equiv H t, \chi \equiv \phi/H, \Pi \equiv {\bar \pi}/H^2$. 
Then,  the basic equations are follows, 
\begin{align}
\frac{\partial \chi}{\partial N} &= \left(1+\frac{3}{\nu-3/2}\right)\Pi (N)+ \frac{1}{(\nu-3/2)}\left(c_{\rm inf}\left(\chi+\frac{\Pi}{\nu-3/2}\right)+\lambda \left(\chi+\frac{\Pi}{\nu-3/2}\right)^3\right), \label{Lange3}\\
\frac{\partial \Pi}{\partial N} &= -3 H \Pi(N)-\left(c_{\rm inf}\left(\chi+\frac{\Pi}{\nu-3/2}\right)+\lambda \left(\chi+\frac{\Pi}{\nu-3/2}\right)^3\right)+{\tilde \tau} (\vect{x},t) , \label{Lange}
\end{align} 
with 
\begin{equation}
\langle 0 | {\tilde \tau}(x_1) {\tilde \tau}(x_2) | 0 \rangle = \frac{\Gamma(\nu)^2}{2^{1-2\nu}} \left|\nu-\frac{3}{2}\right|^2 \frac{1}{4 \pi^3} \delta(N_1-N_2). 
\end{equation}

We solved them by using the Euler-Maruyama method. 
We calculated numerically the following equations, 
\begin{align}
\chi_{n+1} & = \chi_{n} +a_1(\chi_n,\Pi_n)\Delta N \\
\Pi_{n+1} &=\Pi_{n}+a_2(\chi_n,\Pi_n) \Delta N+b(\chi_n,\Pi_n)\Delta W
\end{align}
with 
\begin{align}
a_1(\chi_n,\Pi_n)&=\left(1+\frac{3}{\nu-3/2}\right)\Pi_n+ \frac{1}{(\nu-3/2)}\left(c_{\rm inf}\left(\chi_n+\frac{\Pi_n}{\nu-3/2}\right)+\lambda \left(\chi_n+\frac{\Pi_n}{\nu-3/2}\right)^3\right),  \\
a_2(\chi_n,\Pi_n)&=-3 H \Pi_n-\left(c_{\rm inf}\left(\chi_n+\frac{\Pi_n}{\nu-3/2}\right)+\lambda \left(\chi_n+\frac{\Pi_n}{\nu-3/2}\right)^3\right), \\
b(\chi_n,\Pi_n)&=\frac{\Gamma(\nu)}{2^{(3-2\nu)/2}\pi^{3/2}}\left|\nu-\frac{3}{2}\right|, 
\end{align}
from $N=0$ to 50 (100) with the step width $\Delta N =10^{-3}$
and the initial conditions $\chi=\Pi=\partial \chi/\partial N=\partial \Pi/\partial N=0$. 
Here subscript $n$ represents that the variable is of the $n$-th step, and 
$\Delta W$ is a random variable that satisfies $\langle \Delta W^2\rangle=\Delta N$
generated by the Mersenne-Twister method \cite{MT}. 
We performed $10^6$ trials for each model parameter, $c_{\rm inf}=10^{-2}$ to 1 
(and $\lambda=-0.01$), 
and obtained the result 
shown in Figs.~\ref{fig:hist} and \ref{fig:ave}. 
We stopped calculation once it gets $|{\bar \phi}|/H>30$
since in this case the Higgs field goes down to the AdS vacuum rapidly 
and it will go to infinity. 
We confirmed that the proportion of such trials is less than 2\% even for 
$c_{\rm inf}=10^{-2}$ and ${\cal N}_*=100$. 
Thus it does not affect our result.

\section{The validity of the approximation for the Hubble-induced mass during inflaton oscillation dominated era \label{app}}

In our analytic calculation, we integrate out the inflaton dynamics and treat its effect as the ``Hubble-induced mass'' term in the Higgs potential during inflaton oscillation dominated era. Here, we calculate the time evolution of the Higgs field $h$ numerically without integrating out of the inflaton field $\phi$ in order to demonstrate the validity of our approximation.

We consider the massive chaotic inflation model with a $h^2 \phi^2$ interaction term as a simple example, 
\begin{equation}
V=\frac{1}{2}m^2 \phi^2+\frac{\lambda( h)}{4}h^4+\frac{\kappa}{2}h^2 \phi^2,  
\label{eq:V_app}
\end{equation}
with $m=10^{13}$ GeV. 
For simplicity, we assume an approximate formula, 
\begin{equation}
\lambda(\mu)=-1.4 \times 10^{-3}\ln \left(\frac{\mu}{10^{11} {\rm GeV}}\right), 
\end{equation}
to estimate the scale dependence of the Higgs quartic coupling. From the potential in Eq.(\ref{eq:V_app}), the basic equations are given by 
\begin{align}
&{\ddot h}+3H{\dot h}+\lambda(h) h^3+\frac{\partial \lambda (h)}{\partial h}\frac{h^4}{4}+\kappa \phi^2 h=0, 
\label{eq:app_BE1} \\
&{\ddot \phi}+3H{\dot \phi}+m^2 \phi + \kappa h^2 \phi=0, 
\label{eq:app_BE2} \\
&3H^2M_{\rm Pl}^2=\frac{1}{2}({\dot \phi}^2+{\dot h}^2)+\frac{1}{2}m^2\phi^2+\frac{\lambda(h)}{4}h^4+ \frac{\kappa}{2} h^2\phi^2, 
\label{eq:app_BE3}
\end{align}
where we neglect the spatial derivatives, $\nabla \phi, \nabla h$.

\subsection{Approximate calculation}
First, we estimate the time evolution of the Higgs field with an approximation in which we assume the Higgs field dynamics does not affect on the inflaton dynamics and 
cosmic expansion. In this case, the inflaton oscillation and the Hubble parameter after inflation are given as
\begin{equation}
\phi(t)=2\sqrt{\frac{2}{3}}\frac{M_{\rm Pl}}{mt}\sin(mt), \quad H(t)=\frac{2}{3t}, 
\end{equation}
respectively. By averaging the inflaton oscillation over time, 
\begin{equation}
{\bar \phi}(t)=\frac{2}{\sqrt{3}}\frac{M_{\rm Pl}}{mt}=\sqrt{3}\frac{M_{\rm Pl}}{m}H(t), 
\end{equation}
the ``effective mass'' of the Higgs field can be obtained from the $(\kappa/2) \phi^2 h^2$ coupling as, 
\begin{equation}
m_h^2(t)= \kappa{\bar \phi}(t)^2 = c_{\rm osc} H^2(t),  
\end{equation}
where $c_{\rm osc}=3 \kappa ( M_{\rm Pl}/m )^2$. The averaging of the inflaton oscillation can be justified when the time scale of the inflaton field evolution is much shorter than the Higgs field evolution.

Neglecting the quartic term, the dynamics of the Higgs field after inflation is then described by 
the following equation of motion, 
\begin{equation}
{\ddot h}+\frac{2}{t} {\dot h}+\frac{4c_{\rm osc}}{9t^2} h =0. 
\end{equation}
The solution is given by, 
\begin{equation}
h(t)\simeq h_0 \left(\dfrac{3mt}{2}\right)^{-(1-\sqrt{1-16 c_{\rm osc}/9})/2}. \label{hievo}
\end{equation}
with $h_0$ being the initial condition
for $c_{\rm osc}<9/16$ which we are now interested in. 
The quartic coupling can be neglected only if 
\begin{equation}
\frac{1}{2}m_h^2(t) h^2(t) \gg \frac{|\lambda(h)|}{4} h^4(t) \Leftrightarrow h(t) \ll \sqrt{\frac{2}{|\lambda|}} m_h(t) =2\sqrt{\frac{2\kappa }{3|\lambda|}}\frac{M_{\rm Pl}}{mt},  \label{valid}
\end{equation}
where we neglect the $h$ dependence of $\lambda$. 
If $h(t)$ becomes less than that value, 
the above approximation no longer valid and it rolls down to the 
unwanted true vacuum due to the negative quartic term. 
Combining Eq.~\eqref{hievo} and Eq.~\eqref{valid}, we now have an analytic estimate for the time  
when the Higgs field falls down to the unwanted true vacuum, 
\begin{equation}
mt\simeq \left(\dfrac{2}{3}\right)^{\frac{\sqrt{1-16c_{\rm osc}/9}}{1+\sqrt{1-16c_{\rm osc}/9}}}\left(\dfrac{\kappa}{|\lambda|}\right)^{\frac{1}{1+\sqrt{1-16c_{\rm osc}/9}}} \left(\dfrac{2M_{\rm pl}}{h_0}\right)^{\frac{2}{1+\sqrt{1-16c_{\rm osc}/9}}}. \label{falt}
\end{equation}

\subsection{Numerical calculation}
Next, we calculate the time evolution of the Higgs field numerically. 
Using the following dimensionless variables, 
\begin{equation}
{\hat h}=h/m, \quad {\hat \phi}=\phi/m, \quad {\hat H}=H/m, \quad \eta=mt, 
\end{equation}
Eqs.~(\ref{eq:app_BE1})-(\ref{eq:app_BE3}) are rewritten as
\begin{align}
&\frac{\partial^2 {\hat h}}{\partial \eta^2}+3 {\hat H}\frac{\partial {\hat h}}{\partial \eta}+\lambda({\bar{\hat h}}) {\hat h}^3+\frac{1}{4}\frac{\partial \lambda}{\partial {\hat h}}{\hat h}^4+\kappa {\hat h} {\hat \phi}^2=0,  \\
&\frac{\partial^2 {\hat \phi}}{\partial \eta^2}+3{\hat H}\frac{\partial {\hat \phi}}{\partial \eta}+{\hat \phi} + \kappa {\hat \phi} {\hat h}^2=0,  \\
&3 \left(\dfrac{M_{\rm Pl}}{m}\right)^2{\hat H}^2 =\frac{1}{2}\left(\left(\frac{\partial {\hat \phi}}{\partial \eta}\right)^2+\left(\frac{\partial {\hat h}}{\partial \eta}\right)^2+{\hat \phi}^2 \right)+\frac{\lambda({\bar {\hat h}})}{4}{\hat h}^4+\frac{\kappa}{2}{\hat \phi}^2{\hat h}^2. 
\end{align}

Here,  we investigate the time evolution of this system by using 4-th order Runge-Kutta method with the adaptive step size control taking the initial conditions at $\eta=10^{-4}$ as 
\begin{equation}
{\hat \phi}=2 \frac{M_{\rm Pl}}{m}, \quad \frac{\partial {\hat \phi}}{\partial \eta}=0, \quad {\hat h}={\hat h}_0, \quad \frac{\partial {\hat h}}{\partial \eta}=0, 
\end{equation}
with $h_0=(0.1-1)  \sqrt{\kappa} M_{\rm Pl}$.

\begin{figure}[t]
 \begin{center}
\begin{tabular}{cc}
  \includegraphics[width=85mm]{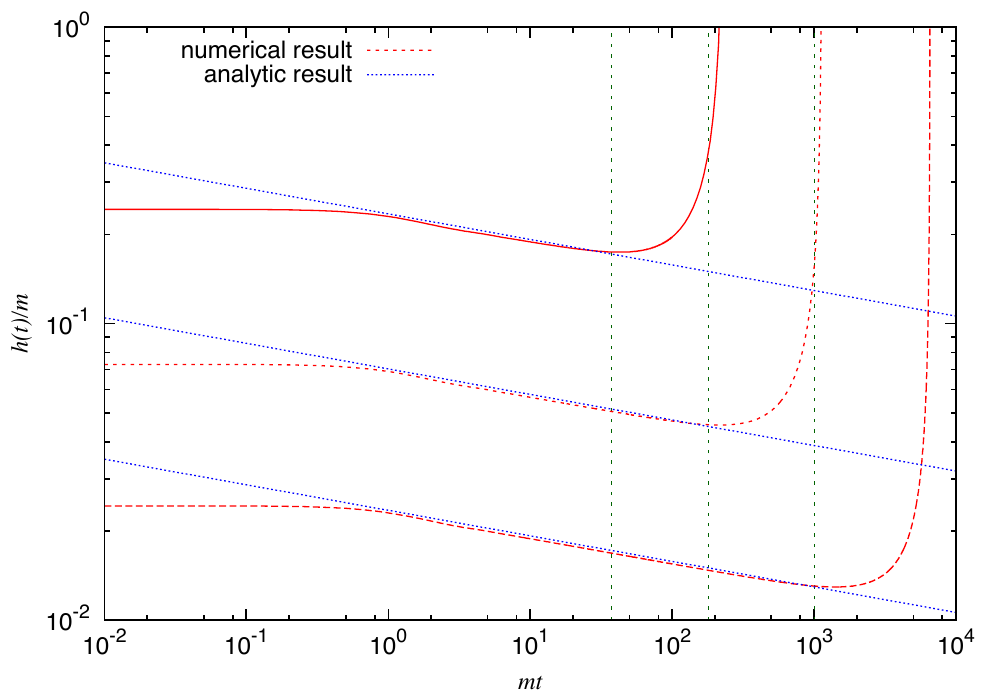}
  &
  \includegraphics[width=85mm]{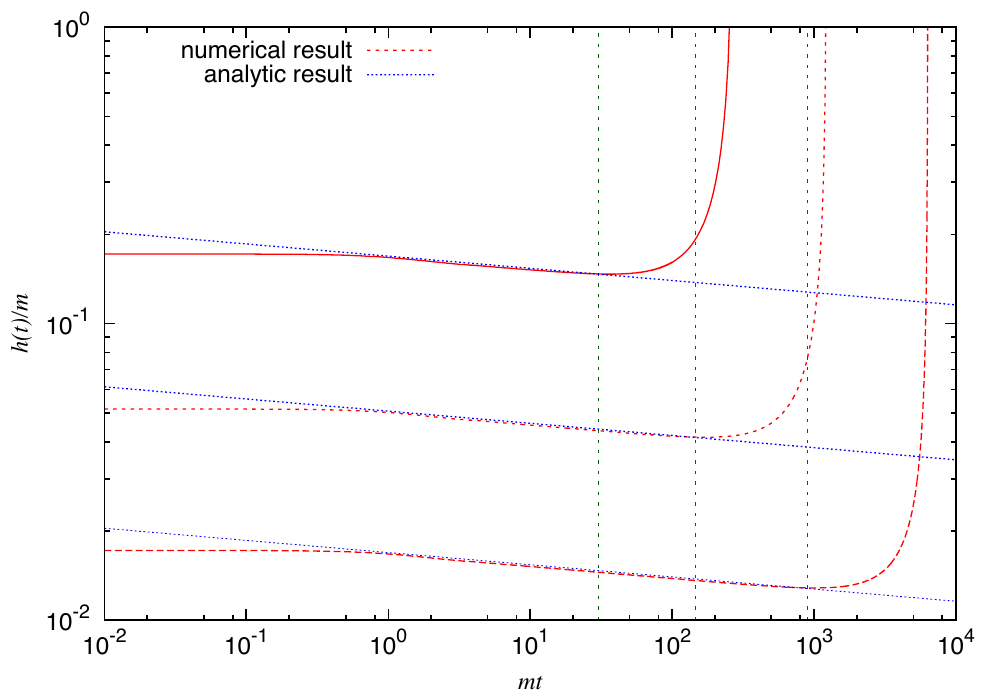}
\end{tabular}
\vspace{-0.3cm}
 \end{center}
 \caption{The time evolution of the Higgs field with the Higgs-portal coupling to inflaton 
 during inflaton oscillation dominated era. 
 Curved red lines represent the numerical result and straight blue lines represent the analytic solution. 
 Vertical green lines represent the analytic estimation of the Higgs fall time (Eq.~\eqref{falt}). 
 Here we take the initial value of the Higgs field as (0.1 (dashed), 0.3(dotted), 1(straight)) 
 $\times \sqrt{\kappa} M_{\rm Pl}$. The Higgs-portal couplings are chosen as $10^{-12}$ (left) and $5 \times 10^{-13}$ (right).  }
 \label{fig:app1} 
\end{figure}

In Fig.~\ref{fig:app1}, we show the numerical results of tracing the Higgs field time evolution (by curved red lines) and the results from approximate calculation, Eq.~\eqref{hievo} (by blue dotted lines) with $\kappa= (1.0 \times 10^{-12}, 5.0\times 10^{-13})$.\footnote{These values correspond to $c_{\rm osc}\simeq (0.177, 0.0886)$.}
The analytic fall times (Eq.~\eqref{falt}) are indicated by vertical green lines. In both results, the initial value of the Higgs field is taken as $h_0/(\sqrt{\kappa} M_{\rm Pl})=0.1$, 0.3 and 1.0 from above. 

The Higgs field starts slow-rolling at $mt=1$ and goes to the true vacuum, finally, when the negative quartic term gets effective. The time when the Higgs field falls into the true vacuum is a little later than the approximated result (Eq.~\eqref{falt}), but this is because it takes time for the Higgs field really to fall down to the unwanted vacuum after it starts to feel the negative quartic term. 
Note that the numerical result starts to deviate from the analytic estimate exactly at the time evaluated in Eq.~\eqref{falt}. 
Thus, our approximate calculation can be useful to estimate the time evolution of the Higgs field approximately as shown in Fig.\ref{fig:app1}.

%%%%%%%%%%%%%%%%%%%%%%%%%%%%%%%%

%%%%%%%%%%%%%%%%%%%%%%%%%%%%%%%%%%

%\newpage

\end{document}